\documentclass[conference, onecolumn]{IEEEtran}
\usepackage[utf8]{inputenc}
\usepackage{mathtools}
\usepackage{amssymb}
\DeclareMathOperator*{\argmax}{argmax}
\usepackage{algorithm}
\usepackage{algpseudocode}
\usepackage{graphicx}
\graphicspath{{./figs/}}
\usepackage{subcaption}
\usepackage[breaklinks=true]{hyperref}
\usepackage[backend=bibtex]{biblatex}
\title{Optimal Seat Allocation Under Social Distancing Constraints}

\author{\IEEEauthorblockN{Michael Barry\IEEEauthorrefmark{1}, Claudio Gambella, Fabio Lorenzi, \\ John Sheehan and Joern Ploennigs} \\
\IEEEauthorblockA{IBM Research Europe \\ 
\IEEEauthorrefmark{1} michael\_barry@ie.ibm.com, ...
}}

\date{\today}

\begin{document}

\maketitle

\begin{abstract}
    The Covid-19 pandemic introduces new challenges and constraints for return to work business planning. We describe a space allocation problem that incorporates social distancing constraints while optimising the number of available safe workspaces in a return to work scenario.
    We propose and demonstrate a graph based approach that solves the optimisation problem via modelling as a bipartite graph of disconnected components over a graph of constraints. We compare results obtained with a constrained random walk and a linear programming approach.

\end{abstract}

\section{Introduction}
The Covid 19 pandemic has raised new challenges in workplace health and safety and is driving new requirements, approaches and innovations in workplace management \cite{ECDC} \cite{OSHA} \cite{IBM}. Respecting social distance between employees in an office environment is a necessary step in providing a safe, comfortable working environment.

Recently there has, understandably, been a lot of interest in solutions that incorporate distance constraints into real-world seat allocation problems. \cite{ActivityStream} and \cite{Softjourn} describe seat allocation solutions for theatres and other entertainment venues that explores the capacity of a venue under different social distancing conditions. An interactive seat selection tool is described in \cite{Robinpowered} that requires the user to manually select individual spaces. A solution for social distancing for public facilities, inspired by previous work on the optimal positions of turbines in offshore wind farms is presented in \cite{MathematicalOptimizationForSocialDistancing}. 

The office environment, presents its own unique challenges to seat allocation and social distance constraints. While an office layout may be more structured than that of a theatre, cinema or other venue, it also comes with a larger variety of space sizes and shapes, as well as other business related operational constraints, including prioritising teams and business units for workspace allocation and the desire to co-locate existing teams so that returning members are not spread across multiple floors or areas. Each organisation brings its own unique use cases, quality metrics, and constraints that must be taken into account in any workspace allocation exercise.

Notwithstanding the workspace allocation problem itself, applying a social distance constraint to every floor in every building of a large organisation is an immense undertaking. Existing building and office layouts must be reviewed and assessed such that only those workspaces that are socially distant from each other can be used. Reviewing, measuring and marking out workspaces is a time consuming task for an space planner or office manager, particularly for large organizations with office workspaces in the thousands. Furthermore, these staff are not data scientists or optimization specialists that are trained to analyse the data and develop an optimal plan. 

We present an automated approach to optimising workspace allocation in office environments based on pre-existing floorplan images. We describe different approaches for identifying workspaces in a floorplan image. Next we propose a graph based approach to optimising workspace allocation using the identified workspaces. Finally we present a Linear Program based approach that incorporates social distance constraints into a general model for workspace allocation in an office environment

\section{Identifying Workspaces} \label{sec:discovery}
Before we can formulate the optimization problem we need to identify workspaces in the floor. Identifying workspaces, or indeed any shape in a floorplan image is a challenging task due to the many formats a floorplan may be represented by. Workspaces may be simple squares for a desk on a floorplan, or a complex 3D model in a CAD file. We use a variety of computer vision techniques and, where available, metadata to identify workspaces in floorplans.

When a floorplan image is a vector file such as a SVG \cite{SVG}, we parse the image to extract Polylines and Paths that describe likely workspaces. We prune this set of likely workspaces based on the size of Before we can formulate the optimization problem we need to identify workspaces in the floor  of the discovered workspaces. This technique works well for office spaces that already make use of regularly shaped and oriented cubicles, where workspaces can be represented as simple polygons. Moreover, it can be adapted to discover more complex objects, such as individual workspaces or desks using a bounding-box that encompasses the element.

Where the floorplan is a raster image, for example a JPEG or PNG file, we us image convolution to detect workspaces in the floorplan \cite{openCV}. By applying a template of a workspace we identify and extract workspaces and their locations in the target image. \autoref{fig:findSeats} below shows the output of a template search on a target floorplan. The workplace templates (a) and (b) are applied to the image and spaces of each type are identified, regardless of orientation or size.

\begin{figure}[htbp]
\centering
  \begin{subfigure}[b]{0.45\textwidth}
    \includegraphics[width=\textwidth]{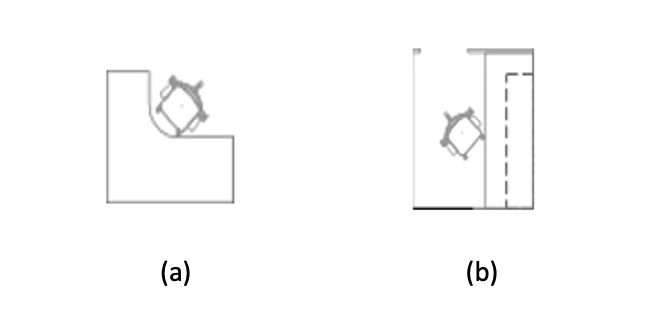}
  \end{subfigure}
  \hspace{1em}%
  \begin{subfigure}[b]{0.45\textwidth}
    \includegraphics[width=\textwidth]{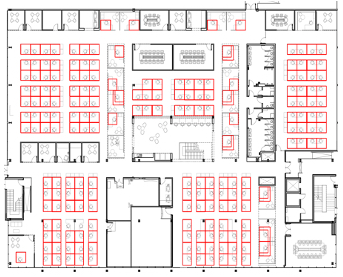}
    \label{subfig:ILP_Floorplan_3}
  \end{subfigure}
  \caption{Using templates to find workspaces in a Floorplan image}
\label{fig:findSeats}
\end{figure}

Metadata from a system of record can also be used to identify workspaces. As well as providing floorplans, commercial space management systems may provide a list of spaces per floor including workspace tags that allows us to clearly identify target workspaces.

\section{Allocating Workspaces} 
Once we have identified a set of workspaces and their location we can develop a seating plan that identifies that maximum set of spaces that can be allocated while maintaining social distancing. The inputs to the space allocation include the desired social distance, and the centroids; the centers of the bounding boxes of the workspaces. The centroids image coordinates are rescaled to human readable (metric or imperial) coordinates using a linear rescaling along the x,y axis.

We developed different approaches to solve the seat assignment problem. The first solution uses a random walk to divide the workspaces into sets of allocated and unallocated workspaces. Next we describe a graphical approach to map the problem to a set of disjoint bipartite graph mapping problems. Finally, we present a Linear Program Formulation that includes a social distancing constraint.

\subsection{Random Walk Heuristic}
The random walk approach selects a workspace at random from the full floorplan and evaluates the distance between the workspace centroid and the centroids of all allocated workspaces. If the workspace is sufficiently distant from all allocated workspaces it is added to the set of allocated workspaces; otherwise it is unallocated. Once all workspaces have been added to the allocated or unallocated sets, the set of allocated workspaces is returned.

This is a naive approach and does not result in an optimal allocation of workspaces. The accuracy of the solution can be improved by running the heuristic multiple times and selecting the individual solution with the maximum allocation of workspaces.  

\subsection{Graph Analysis}
\label{sec:GraphAnalysis}
In graph analysis, a selection problem is a problem that involves splitting a set of resources $R$ into two disjoint sets $S$ of selected resources and $N$ of non-selected resources such that $S\cap N=\emptyset$ and $R=S\cup N$ according to some selection criterion (or function) while maximizing the cardinality of the selected nodes' set $\vert S\vert $. In cases where the selection function is dependent only on the elements of $R$ themselves $f\colon  R\to \{0,1\}$ the problem can be treated as a graph problem. 

\begin{figure*}[htbp]
\centering
  \begin{subfigure}[b]{0.3\textwidth}
    \includegraphics[width=\textwidth]{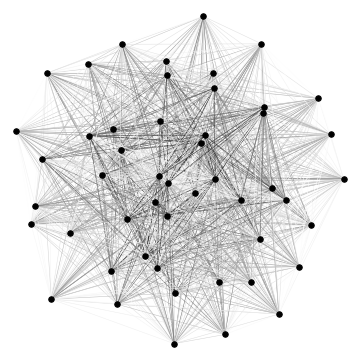}
    \caption{$K_{50}$}
    \label{subfig:k50}
  \end{subfigure}
  \hspace{1em}%
  \begin{subfigure}[b]{0.31\textwidth}
    \includegraphics[width=\textwidth]{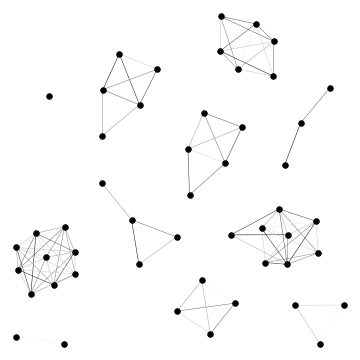}
    \caption{Graph of contstraints}
    \label{subfig:constraints}
  \end{subfigure}
  \hspace{1em}%
  \begin{subfigure}[b]{0.3\textwidth}
    \includegraphics[width=\textwidth]{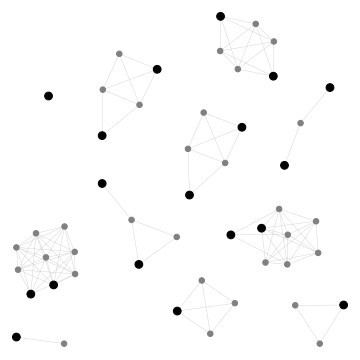}
    \caption{Seat selection}
    \label{subfig:selection}
  \end{subfigure}
  \caption{Visualisation of the three stages of the seat selection algorithm}
  \label{fig:algo}
\end{figure*}

\subsubsection{Graph of Constraint}
We construct a weighted graph where each node is a workspace from the floorplan and the edges between nodes are weighted using the distances between workspace centroids. This results in a densely connected graph as shown in \autoref{subfig:k50} where each node is connected to every other node in the graph. This large, fully connected graph is unwieldy, and includes many edges whose weight likely outside the social distance that we wish to apply. To reduce the size and complexity of the graph we define a ``graph of constraint`` as shown in \autoref{subfig:constraints} that only includes those nodes connected by edges with weights that are within social distance.

In the workspace allocation scenario under social-distancing constraints $R$ is the set of available workspaces. Let $d$ be the minimum required social distance. Let $K_{\mid R\mid }=\{R,E,w\}$ be a complete undirected weighted graph. $E=(R\times R)$ be the set of edges of the graph.  $w\colon  E\to\mathbb{R}$ be a weight function that assigns each edge a weight based on the distance between the spaces on the floorplan. Let $C_d=\{R',E',w\}$ be a subgraph of $K_{\mid R\mid }$ where $R'\subset R$ and $E'=\{e_i\mid e_i\in E, w(e_i)<k\}$. $C_d$ is a potentially disconnected graph named the ``graph of constraint``.

\subsubsection{Allocating workspaces using a Bipartite Graph}
A bipartite graph is one whose nodes can be divided into two disjoint, independent sets such that a node in one set can only connect to nodes in the other set, and not to nodes in its own set\cite{Salvatore}. If the ``graph of constraints`` is a bipartite graph, or can be made bipartite, then the two sets of nodes represent independent sets of allocatable workspaces. As the nodes in each set are not connected to each other, they are outside social distance of each other. We then choose the larger set as the allocated workspaces for the floor, and all other workspaces are unallocated.

Let $B_d\subset C_d$ be a bipartite subgraph of $C_d$. We denote the set of nodes of a graph $G$ with $N(G)$ and the set of edges with $E(G)$.
Given a set of workspaces $R$ and a minimum social distance parameter $d$ we can obtain an acceptable selection of workspaces by partitioning the graph and selecting the partition with the most nodes as described in Algorithm~\ref{alg:seat_selection}. A visualization of the algorithm is shown in \autoref{fig:algo}.

\begin{algorithm}[htbp]
    \caption{Workspace selection algorithm}
    \label{alg:seat_selection}
    \begin{algorithmic}[1] 
        \Procedure{SpaceSelection}{$C_d$} \Comment{The graph of constraints $C_d$}
            \State $S\gets\emptyset$
            \ForAll{$C'\in C_d$} \Comment{We check each connected component of $C_d$}
                \State $B_d \gets partition(C')$
                \State $U, V \gets bicolor(B_d)$
                \State $S' \gets\argmax_{X\in\{U,V\}}(\mid X\mid )$ \Comment{We select the partition with maximum cardinality}
                \State $S\gets S\cup S'$
            \EndFor
            \State \textbf{return} $S$\Comment{A selection of workspaces satisfying the spatial constraints on $d$}
        \EndProcedure
    \end{algorithmic}
\end{algorithm}

If the graph is not a bipartite graph, we delete nodes from the graph until the remaining graph is bipartite. After a node is deleted from the graph it is no longer available to become allocated in any remaining bipartite graph, and so that workspace must become unavailable on the floorplan. When deleting nodes from the graph, we must ensure that a maximum number of nodes remains in the bipartite graph, so that a maximum number of workspaces are available for allocation.

\begin{figure*}[htbp]
\centering
  \begin{subfigure}[b]{0.2\textwidth}
    \includegraphics[width=\textwidth]{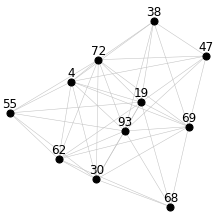}
    \caption{$C'$}
    \label{subfig:component}
  \end{subfigure}   
  \hspace{1em}%
  \begin{subfigure}[b]{0.2\textwidth}
    \includegraphics[width=\textwidth]{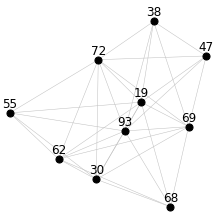}
    \caption{$c=4$}
    \label{subfig:h1}
  \end{subfigure}
  \hspace{1em}%
  \begin{subfigure}[b]{0.2\textwidth}
    \includegraphics[width=\textwidth]{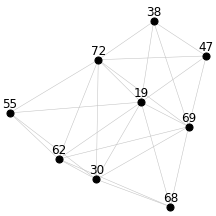}
    \caption{$c=93$}
    \label{subfig:h2}
  \end{subfigure}
  \hspace{1em}%
  \begin{subfigure}[b]{0.2\textwidth}
    \includegraphics[width=\textwidth]{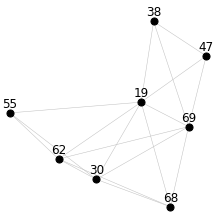}
    \caption{$c=72$}
    \label{subfig:h3}
  \end{subfigure}
  \begin{subfigure}[b]{0.2\textwidth}
    \includegraphics[width=\textwidth]{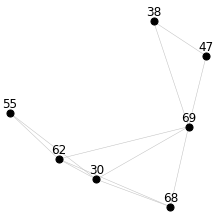}
    \caption{$c=19$}
    \label{subfig:h4}
  \end{subfigure}
  \hspace{1em}%
  \begin{subfigure}[b]{0.2\textwidth}
    \includegraphics[width=\textwidth]{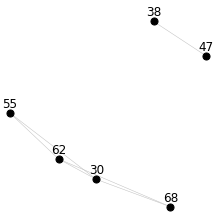}
    \caption{$c=69$}
    \label{subfig:h5}
  \end{subfigure}
  \hspace{1em}%
  \begin{subfigure}[b]{0.2\textwidth}
    \includegraphics[width=\textwidth]{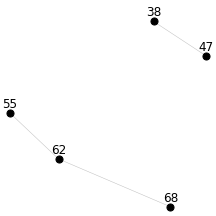}
    \caption{$c=30$}
    \label{subfig:h6}
  \end{subfigure}
  \hspace{1em}%
  \begin{subfigure}[b]{0.2\textwidth}
    \includegraphics[width=\textwidth]{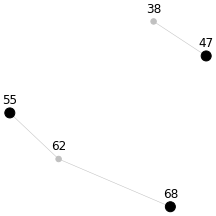}
    \caption{bipartition}
    \label{subfig:bipartition}
  \end{subfigure}
  \caption{A visualisation of the heuristic candidate selection algorithm. The sequence of images from (\subref{subfig:component}) to (\subref{subfig:bipartition}) represents the stages of the selection and elimination of the nodes}
  \label{fig:heur_selection}
\end{figure*}

\autoref{fig:heur_selection} depicts how nodes are selected for deletion. Examining \autoref{subfig:h5}, the depicted graph is not bipartite as it contains odd cycles: {68-30-62-68} and {55-62-30-55}. Nodes 62 and 30 are present in both odd-cycles and so are candidates for elimination. Each candidate has three outgoing connections, so one is chosen at random for deletion. Consider that for node 30 to be in the set of allocated workspaces, nodes 68, 62 and 55 cannot be allocated; they are connected to node 30 and so are within social distance of node 30 and must remain unallocated. This would reduce the number of nodes in the graph to three: nodes 30, 38 and 47. Deleting node 30 results in a larger graph with five nodes. Node 30 is deleted and the resulting graph is shown in \autoref{subfig:h6}. This graph is bipartite and can be partitioned as shown in \autoref{subfig:bipartition}. 

The heuristic to transform an odd cycle graph to a bipartite graph through through node elimination is described in Algorithm~\ref{alg:partition} and Algorithm~\ref{alg:selecth}. As a graph with odd-cycles cannot be bipartite, the heuristic first examines the cycle basis of the graph to identify nodes that participate in the most number of odd-cycles in the graph. Then the node with the most outgoing connections is selected for elimination. Nodes are eliminated until the graph contains no odd-cycles and is bipartite.

\begin{algorithm}[htbp]
    \caption{Partitioning algorithm}
    \label{alg:partition}
    \begin{algorithmic}[1] 
        \Procedure{Partition}{$C'$} \Comment{A component of the graph of constraints $C_d$}
            \State $X\gets cycle\_basis(C')$
            \State $X'\gets\{x\mid x\in X, \mid N(x)\mid \bmod 2\neq 0\}$
            \While{$X'\neq\emptyset$}
            \State $c\gets candidateH(X')$\Comment{Apply the heuristic-based candidate selection and select a candidate node for elimination}
            \State $N(C')\gets N(C')\setminus \{c\}$
            \State $E(C')\gets E(C')\setminus \{e\mid e\in E(C'), c\notin e\}$
            \State $X'\gets \{x\mid c\notin N(x), x\in X'\}$
            \EndWhile
            \State \textbf{return} $C'$ \Comment{Return $C'$ without odd cycles}
        \EndProcedure
    \end{algorithmic}
\end{algorithm}

\begin{algorithm}[htbp]
    \caption{Heuristic candidate selection algorithm}
    \label{alg:selecth}
    \begin{algorithmic}[1] 
        \Procedure{CandidateH}{$X'$} \Comment{A cycle basis for $C'$}
            \State $C\gets \argmax_{x\in\bigcup_{y\in X'}N(y)}\sum_{X_i\in X'}[x\in X_i]$ \Comment{Select the candidate with the highest participation in the cycle basis}
            \State $C\gets \argmax_{x\in C}(deg(x))$ \Comment{Select the candidate with the highest out degree}
            \State \textbf{return} $random(C)$ \Comment{Return one element from the set C}
        \EndProcedure
    \end{algorithmic}
\end{algorithm}

\subsection{Seat Allocation using Integer Linear Programming}
The heuristic graph partitioning approach presented before is very efficient in solving simple assignment problems, but is limited in encoding more complex constraints.Each return to the office scenario is unique and must take into account business requirements as well as social distancing. This may, for example, necessitate prioritising some business units or teams over others, or ensuring that teams that do return to the office are seated together and not scattered across multiple floors or buildings. In this section we integrate social distancing into a simplified allocation model for an office environment formulated as an Integer Linear Programming model. The aim of the model is to maximize the number of workspaces allocated across a set of business units, while respecting constraints of a feasible allocation plan. For the sake of simplicity, we focus on the workspaces allocation of a single floor.

First, we describe a mathematical formulation that provides a set of basic requirements for workspace allocation. We denote with $e_j$ the number of employees of business unit $j$, $c_{ij}$. We introduce binary variables $x_{ij}$, to indicate whether the workspace $i$ is allocated to business unit $j$. Determining the optimal seat allocation means solving the following Integer Linear Programming formulation:

\begin{subequations}
	\label{main-prob}
\begin{eqnarray}
  \max_x \sum_{ij} x_{ij} & \label{obj}\\
  \sum_j x_{ij} \leq 1 && \forall\ i \label{con:seat}\\
  \sum_i x_{ij} \leq e_j && \forall\ j \label{con:empl}
\end{eqnarray}
\end{subequations}

The objective \eqref{obj} is to maximimse the number of allocated workspaces. Note that a workspace is not allocated to more than 1 business unit, because of constraints \eqref{con:seat}. Constraints \eqref{con:empl} allocates no more workspaces to a business unit than the number of its employees. Being a generalized assignment problem, Model \eqref{obj}-\eqref{con:empl} can be efficiently solved with commercial optimization solvers. We have adopted CPLEX \cite{CPLEX} for this purpose.

We accommodate social distance in the model by introducing a constraint based on the distance between workspaces, which we can learn by analysing a floorplan as described above. Assuming the distance $d_{ik}$ between each pair $(i,k)$ of workspaces is known, we denote with $I = \{(i, k): d_{ik}< D, i<k \}$ a set of workspace pairs which cannot both be occupied simultaneously. Since the distance requirement between workspaces does not pertain to which business unit they belong to, we introduce auxiliary binary variables $y_i, \forall$ workspace $i$ to indicate whether a workspace is assigned to any business unit. Therefore, social distancing is introduced by adding

\begin{eqnarray}
y_i = \sum_j x_{ij} &  \forall\ i	\label{con:def-y}\\
y_i + y_k \leq 1 & \forall\ (i,k) \in I \label{con:soc-distance}
\end{eqnarray}

to Model \eqref{obj}-\eqref{con:empl}. While constraints \eqref{con:def-y} define whether each workspace is allocated, constraints \eqref{con:soc-distance} enforce the social distance between allocated workspace. Note that variables $y_i$ are bounded by $1$, because of constraints \eqref{con:seat}.

\subsubsection{"Preserving an existing allocation plan"}
It is likely that in an office environment we are working with an existing allocation plan. In general, we wish to minimize variations from that plan so that where possible business units maintain existing  allocations, albeit with less overall workspaces. We discourage changes in the allocation plan $\tilde{x}$ in the objective function \eqref{obj} via penalization terms. Considering variables $x_{ij}$ are binary, it is possible to penalize the variations from allocation plan $\tilde{x}$ simply adding the following linear terms to the objective \eqref{obj}: 

\begin{align}
 C \cdot x_{ij} \quad \forall i \in \tilde{I}, j \in \tilde{J} \quad \text{if} \;  \tilde{x}_{ij}=1 \\
  - C \cdot x_{ij} \quad \forall i \in \tilde{I}, j \in \tilde{J} \quad \text{if} \; \tilde{x}_{ij}=0,
\end{align}

where $C>0$ is an appropriately large constant that acts a penalization factor for the variations from $\tilde{x}$. The larger $C$ is, the more likely CPLEX is to discourage variations from the given allocation plan $\tilde{x}$.

\begin{figure}[htbp]
\centering
  \begin{subfigure}[b]{0.45\textwidth}
    \includegraphics[width=\textwidth]{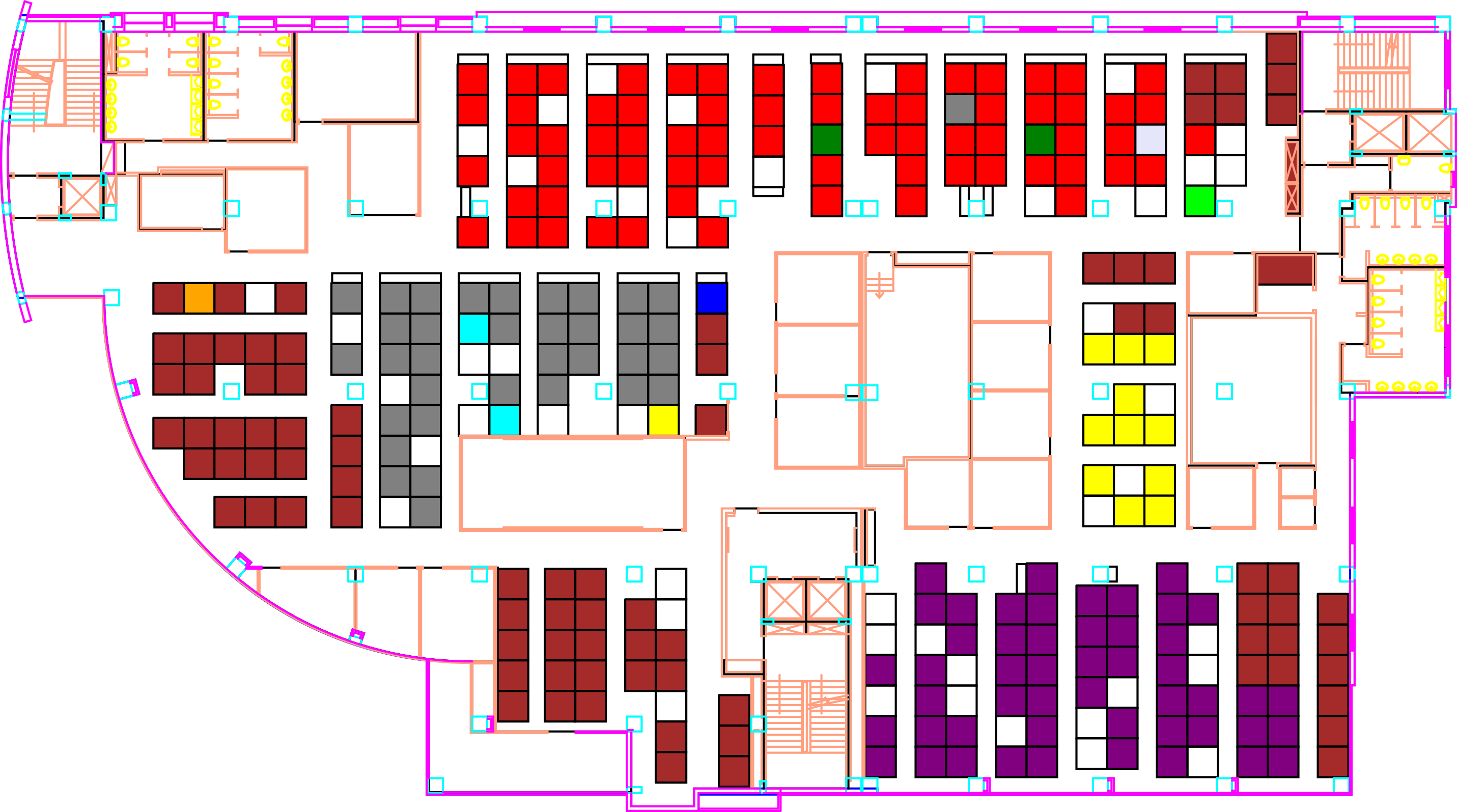}
    \caption{$Existing Floorplan$}
    \label{subfig:ILP_Floorplan_1}
  \end{subfigure}
  \hspace{1em}%
  \begin{subfigure}[b]{0.45\textwidth}
    \includegraphics[width=\textwidth]{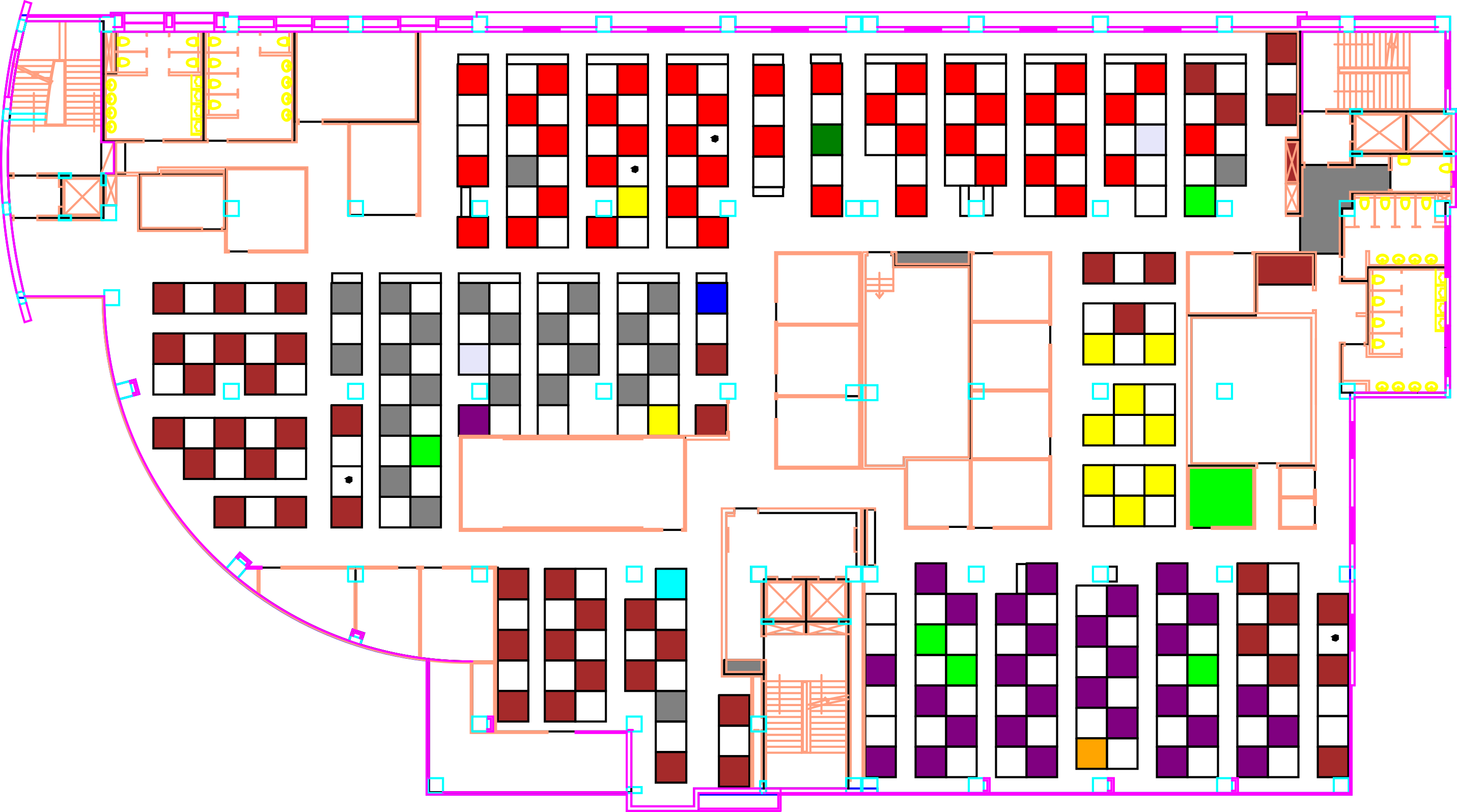}
    \caption{$Social Distance Floorplan$}
    \label{subfig:ILP_Floorplan_2}
  \end{subfigure}
  \caption{Space Allocation with Social Distance}
  \label{fig:ILP}
\end{figure}

The result of a seat allocation using an existing floorplan are shown in \autoref{fig:ILP}. Different colors in the floorplan represent workspaces allocated to different business units in an initial allocation plan. After a social distance is applied,with few exceptions, existing workspace allocations are reused by the same business units.

\section{Results}
\begin{figure}[htbp]
\centering
  \begin{subfigure}[b]{0.45\textwidth}
    \centerline{\includegraphics[width=\textwidth]{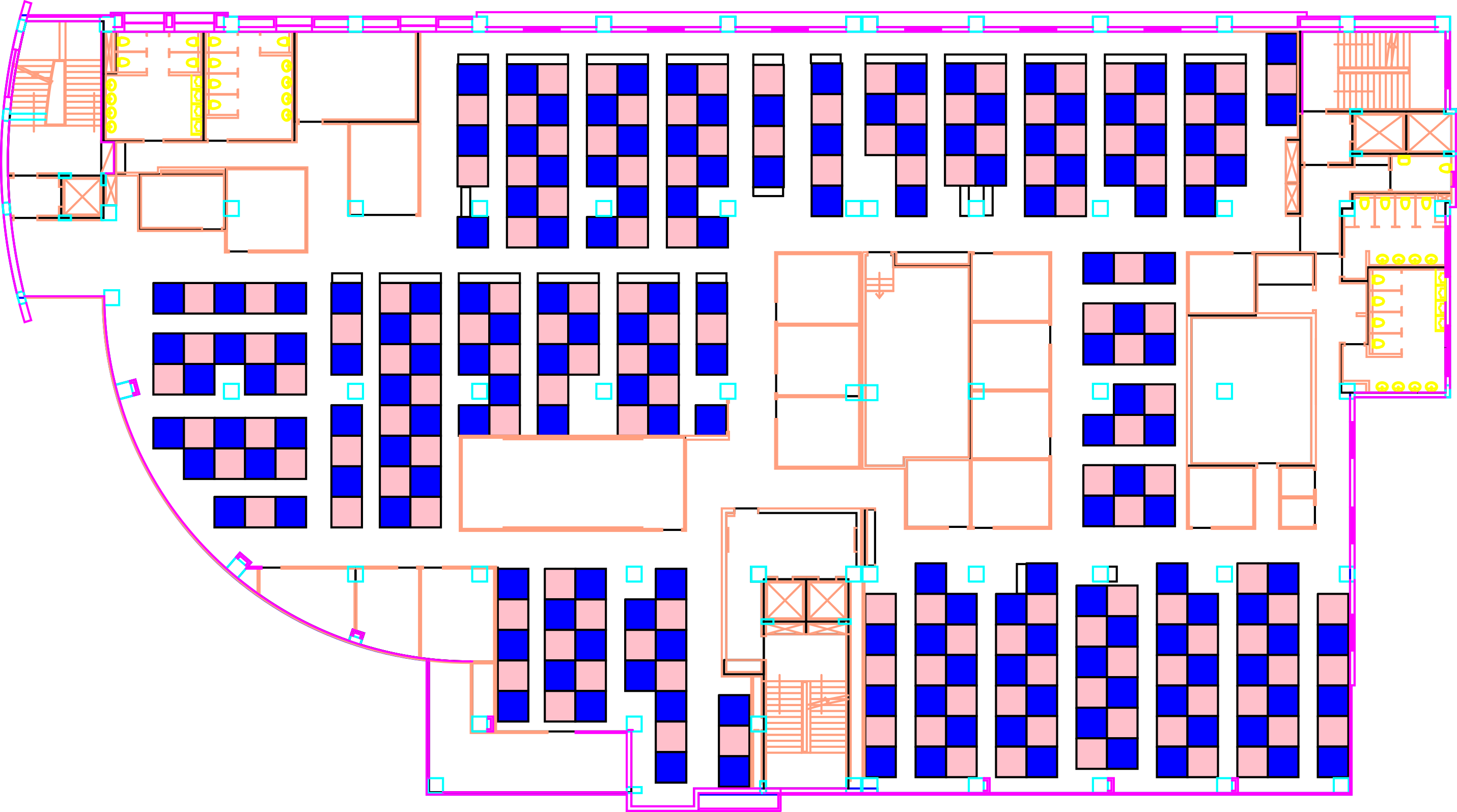}}
  \end{subfigure}
    \caption{Allocating workspaces via graph partition. Allocated workspaces are blue, unallocated are pink}
    \label{fig:SelectedSeats}
\end{figure}

The result of a workspace allocation by creating a bipartite graph  from a floorplan is illustrated in \autoref{fig:SelectedSeats}.The input floorplan is an SVG file with 300 workspaces; discovered from parsing the file as described in \autoref{sec:discovery}. Allocated workspaces are blue, unallocated workspaces are pink. The social distance factor is 72 inches. Workspace dimensions are 60 inches x 60 inches. At a social distance of 72 inches horizontally adjacent and vertically adjacent workspaces are not both allocated as the distance between those workspace centroids is less than the social distance of 72 inches. 

For the same floorplan, he Linear Program and bipartitie graph approaches both identify a maximum of 162 workspaces out of a total of 300 that can be allocated at this social distance. The Random walk approach identifies a maximum allocation of 159 workspaces

 \autoref{tab:SocialDistance} compares the maximum number of allocated desks for the floorplan for a range of social distances and allocation methods. The random walk approach yields the fewest number of allocatable desks. The graph partition and Linear Program produce similar results for distances of 72 inches and 84 inches. At higher distances the Linear Program outperforms the graph partition approach in identifying a maximum number of workspaces that can be allocated

\begin{table}[htbp]
\centering
 \begin{tabular}{|c |c c c|} 
 \hline
 &\multicolumn{3}{c|} {Allocation Approach} \\ \cline{2-4}
 Social Distance (inches) & Random Walk & Partition &  ILP \\ 
 \hline\hline
 72 & 149 & 162& 162 \\ 
 \hline
 84 & 146 & 162& 162 \\
 \hline
 96 & 114 & 123& 123 \\
 \hline
 108 & 99 & 100& 107 \\
 \hline
 \end{tabular}
 \caption{\label{tab:SocialDistance} Allocated Workspaces at a range of social distances for a single workspace}
\end{table}

\autoref{tab:SocialDistanceLargeTable} compares the maximum number of allocated workspaces for a range of floorplans. Each workspace has a different number of workspaces ranging from 173, to 653. The size of the workspaces varies from floor to floor, as does the arrangement of workspaces on each floor. The 1st column contains the number of workspaces on each floor. Subsequent columns contain the maximum number of seats that can be allocated using different approaches at a given social distance. 

As expected, the random walk approach produces the worst results and identifies the least number of seats that can be allocated. The graph partition and Linear Program produce similar results for distances of 72 inches and 84 inches.

At higher distances the results from graph partition approach does not perform as well as the Linear Program. A higher social distance implies that the workspace allocation is more constrained and that there will be more connected nodes in the floorplan ``graph of constraints``. This requires that more nodes be deleted before the graph becomes bipartite. If the heuristic is too aggressive in deleting nodes then an optimal allocation of nodes will not be acheived.

\begin{table}[htbp]
\centering
    \begin{tabular}{|c|cccccccccccc|}
    \hline
     &&\multicolumn{3}{c}{Random Walk}    &&    \multicolumn{3}{c}{Bi-Partition} &&  \multicolumn{3}{c|}{CPLEX} \\
     Workspaces && 72& 84& 96&&  72& 84& 96&& 72& 84& 96  \\
     \hline
     \hline
     173&&  81&  80&  51&&  85&  85& 51&&  85&  85& 51  \\
     267&& 120& 124&  82&& 130& 129& 82&&  130& 129& 86 \\
     300&& 149& 146&  99&& 162& 162& 104&& 162& 162& 109 \\ 
     309&& 158& 157& 114&& 167& 167& 123&& 167& 167& 123  \\
     510&& 238& 238& 134&& 255& 255& 134&& 255& 255& 134  \\
     564&& 237& 163& 159&& 288& 166& 164&& 288& 166& 165  \\
     653&& 304& 185& 146&& 334& 199& 161&& 334& 205& 185  \\
\hline
\end{tabular}
 \caption{\label{tab:SocialDistanceLargeTable} Allocatable Workspaces at a range of social distances for different workspaces}
\end{table}

\section{Conclusion}
Enforcing social distances in office environments is a necessary step in facilitating return to work planning. Automation of floorplan analysis and workspace allocation can greatly improve the speed and efficiency of selecting workspaces that meet social distance constraints. This paper describes an approach that combines image processing techniques with optimization approachess to solve this problem.
This paper demonstrates the use of standard image processing techniques to extract workspace data from floorplan images. From the workspace data, we build a connected graph of workspaces, and reduce that graph to a bipartitionable graph of workspaces by applying social distance constraints. We further illustrate how social distance constraints can be incorporated more complex decision support systems through a linear programming formulation for workspace allocation that incorporates enterprise constraints as well as social distance constraints. 

\printbibliography
\end{document}